\documentclass{article}

\usepackage{arxiv}

\usepackage[utf8]{inputenc} 
\usepackage[T1]{fontenc}    
\usepackage{hyperref}       
\usepackage{url}            
\usepackage{booktabs}       
\usepackage{amsfonts}       
\usepackage{nicefrac}       
\usepackage{microtype}      
\usepackage{lipsum}
\usepackage{graphicx}
\usepackage{amsmath}

\usepackage[section]{placeins}

\title{Construction of embedded fMRI resting state functional connectivity networks using manifold learning}

\author{
 Ioannis K. Gallos \\
  School of Applied Mathematical and Physical Sciences\\
  National Technical University of Athens, Greece\\
  \texttt{yiannis.gallos@gmail.com} \\
   \And
 Evangelos Galaris \\
  Dipartimento di Matematica e Applicazioni\\
  Universit\`a degli Studi di Napoli Federico II \\
  \texttt{evangelos.galaris@unina.it} \\
  \And
   Constantinos I. Siettos \thanks{Corresponding author: constantinos.siettos@unina.it} \\
     Dipartimento di Matematica e Applicazioni \\
     Universit\`a degli Studi di Napoli Federico II \\
    Tel.: +39 0816-75615 \\
    \texttt{constantinos.siettos@unina.it} \\
}

\begin{document}

\maketitle

\begin{abstract}
We construct embedded functional connectivity networks (FCN) from benchmark resting-state functional magnetic resonance imaging (rsfMRI) data acquired from patients with schizophrenia and healthy controls based on linear and nonlinear  manifold learning algorithms, namely, Multidimensional Scaling (MDS), Isometric Feature Mapping (ISOMAP) and Diffusion Maps. Furthermore, based on key global graph-theoretical properties of the embedded FCN, we compare their classification potential using machine learning techniques. We also assess the performance of two metrics that are widely used for the construction of FCN from fMRI, namely the Euclidean distance and the lagged cross-correlation metric. We show that the FCN constructed with Diffusion Maps and the lagged cross-correlation metric outperform the other combinations.

\end{abstract}


\section{Introduction}
\label{intro}

Over the past years, functional magnetic resonance imaging (fMRI) has been widely used for the identification of brain regions that are related to both functional segregation and integration. Regarding functional segregation, the conventional analysis relies on the identification of the activated voxels based on functional response models and multivariate statistics between experimental conditions (e.g. resting state vs. task-stimulated activity). A representative example is the General Linear Model (GLM) that is implemented in well established software packages such as SPM \cite{friston1994statistical} and FSL \cite{smith2004advances}. On the other hand, for the assessment of functional integration, there is a distinction between functional and effective connectivity \cite{friston2011functional}. Functional connectivity (FC) analysis seeks for statistical dependencies (e.g. correlations, coherence) between  brain regions. Effective connectivity (EC) analysis \cite{friston2011functional} tries to reveal the influence that one neural system exerts on another  \cite{friston2011functional}. A detailed review on the differences between  FC and EC approaches can be found in \cite{friston2011functional}.\\
Here, we focus on the construction of FCN based on resting-state fMRI (rsfMRI) recordings. In rsfMRI, there is no stimuli and thus the assessment of functional integration is more complex and not so straightforward compared to task-related experiments \cite{khosla2019machine}. Furthermore, spontaneous/ resting state brain activity as measured with fMRI has been also considered as a potential biomarker in psychiatric disorders (see e.g. the review of Zhou et al. \cite{zhou2010spontaneous}). In general, for the construction of FCN, two basic general frameworks are explored: (a) Seed-based Analysis (SBA) and (b) Independent Component-based Analysis (ICA). In the SBA \cite{biswal1995functional}, the (averaged) fMRI signals of the regions of interest (ROI) are correlated with each other; correlations above a threshold are considered functional connections between seeds/ROIs. Even though the SBA has been proved extremely useful in identifying functional networks of specific brain regions \cite{greicius2003functional,fox2005human,margulies2007mapping}, its major disadvantage is the requirement of the a-priori knowledge of the functional organization of the brain, while possible correlations between seeds can be due to structured spatial confounds (e.g. scanner artifacts) \cite{cole2010advances}. Furthermore, seed selection is based on standard coordinates, while at the subject level, anatomical differences  may lead to the selection of functionally irrelevant voxels at the group level. Thus, despite the use of normalization techniques, the accuracy of this approach is shown to be limited \cite{saxe2006ablation}.
On the other hand, ICA \cite{hyvarinen2000independent} has arisen as an alternative/complementary approach since the early 2000s  \cite{beckmann2005investigations,beckmann2005tensorial,Kim_2010}. ICA decomposes the 4D fMRI data to a set of spatial components with maximum statistical independence and their associated time series. Smith et al.\cite{smith2009correspondence} in a meta-analytic study of 30,000 rsfmRI scans with the aid of ICA revealed a functional ``partition" of the brain into resting-state Networks (RSNs), such as the Sensorimotor, Default mode and Auditory networks. Applications of ICA include also data pre-processing, where noise-related components are regressed out from the original fMRI signals \cite{pruim2015ica}. However, while ICA produces spatial components that are statistically independent to each other, there is no clear link between the spatial components and specific brain functions and furthermore the spatial components cannot in general be ordered by relative importance \cite{cole2010advances}. Another issue is that most of the standard algorithms that compute ICs utilize gradient based optimization algorithms that use an iterative scheme; the initial guesses in these algorithms are generated randomly making the whole process stochastic: for the same dataset, the obtained spatial components may differ significantly over repeated runs \cite{himberg2004validating}. Thus, the robustness/reproducibility of the ICA results over repeated runs may be questioned.\\
In order to tackle the above issues, several techniques have been proposed for the classification of ICs and the construction of subject specific ROIs \cite{pamplona2020personode,yang2008ranking}. Advances have been also made regarding the selection of the model order of the ICA decomposition,  such as the Bayesian dimensionality estimation technique \cite{beckmann2005investigations} and the use of theoretical information criteria for model order selection \cite{li2007estimating}. Finally, the so-called ranking and averaging ICA by reproducibility (RAICAR) \cite{yang2008ranking,himberg2004validating} (see also \cite{cole2010advances} for a critical discussion) aims at resolving issues regarding  stochasticity and robustness of the ICA decomposition. RAICAR utilizes a sufficient number of ICA realizations and based on the reproducibility of the ICs aims to rank them in terms of the most ``reliable" components. Reliable ICs among realizations are assessed via correlations and the final estimate of each component is averaged.\\
Alternatively and/or complementary to the above analysis, linear manifold learning algorithms such as Principal Component Analysis (PCA) \cite{jollifeprincipal,worsley2005comparing,baumgartner2000comparison} and Multidimensional Scaling (MDS) \cite{kruskal1964multidimensional,friston1996functional} have been  exploited. 
PCA has been succesfully applied in the pre-processing routine for dimensionality reduction (often prior to ICA) \cite{iraji2016connectivity}. Applications of PCA, include also the recovery of signals of interest
\cite{viviani2005functional} and the construction of FCN from fMRI scans in task-related experiments   \cite{worsley2005comparing,baumgartner2000comparison}. In these studies, the performance of PCA with respect to the detection of regions of correlated voxels has been shown to be satisfactory but not without problems. For example, a study by Baumgartner et al. \cite{baumgartner2000comparison} highlighted the limits of PCA to correctly identify activation of brain regions in cases of low contrast-to-noise ratios (CNR) appearing  when signal sources of e.g. physiological noise are present \cite{li2009review}.\\
MDS has been also widely used in fMRI (mostly for task-based studies) mainly for the identification of similarities between brain regions in terms of voxel-wise connectivity \cite{shinkareva2013examining,tzagarakis2009cerebral,o2007theoretical,haxby2001distributed,shinkareva2012exploring,de2010distributed}. The implementation of MDS in neuroimaging dates back to the work of Friston et al. \cite{friston1996functional}. There, it was investigated the embedded (voxel-wise) connectivity of fmRI data acquired during tasks of word generation between healthy and schizophrenia subjects. Salvador et al. \cite{salvador2005neurophysiological} used MDS to investigate the embedded connectivity of anatomical regions of the brain from rsfMRI data. Benjaminsson et al. \cite{benjaminsson2010novel} used MDS to embed high-dimensional rsfMRI data from the mutual information space to a low dimensional euclidean space for the identification of RSNs. Herve et al. \cite{herve2012disentangling} used MDS to acquire a low dimensional approximation of interregional correlations for the investigation of the affective speech comprehension. Finally, in a meta-analytic study by Etkin et al. \cite{etkin2007functional}, MDS was exploited to provide a low-dimensional visualization of co-activation interrelations of ROIs. MDS has been also used in works investigating the  functional (dys)connectivity associated with schizophrenia \cite{welchew2002multidimensional} and Asperger’s Syndrome \cite{welchew2005functional}.\\
However, thus far, only a few studies have exploited non-linear manifold learning algorithms such as Local Linear Embedding (LLE) \cite{roweis2000nonlinear}, Isometric Feature Mapping (ISOMAP) \cite{tenenbaum2000global} and Diffusion Maps \cite{coifman2006diffusion} for the analysis of fMRI data and particularly  for the construction of FCN. LLE has been applied in rsfMRI studies for the improvement of predictions in ageing studies \cite{qiu2015manifold} for low-dimensional clustering towards the classification of healthy subjects and patients with schizophrenia \cite{shen2010discriminative} and as an alternative method for dimensionality reduction before the application of ICA in task-related fMRI where non-linear relationships in the BOLD signal are introduced  \cite{mannfolk2010dimensionality}.\\
In Anderson et al. \cite{anderson2013decreased}, ISOMAP was employed to a benchmark rsfMRI dataset of 146 subjects for the construction of embedded low-dimensional FCN for the classification between controls and schizophrenia patients. ROIs were selected using single-subject ICA and the similarities between the ICA components were assessed using a pseudo-distance measure based on lagged cross-correlation. Graph-theoretical measures were then used for classification between patients and healthy controls.
Another study based on single-subject ICA, exploited ISOMAP to classify spatially unaligned fMRI scans \cite{anderson2010classification}. The study considered patients with schizophrenia versus healthy controls and different age groups (young/old) of healthy controls versus patients with alzheimer's disease. Despite the relatively low sample sizes, results were promising with good classification rates. Recently, Haak et al. \cite{haak2018connectopic} utilized ISOMAP in an effort to create individualised connectopies from rsfMRI recordings taken from the WU-Minn Human Connectome Project in a fully data-driven manner.\\
Thus, only a handful of studies have used Diffusion Maps for the analysis of fMRI data, focused on the clustering of spatial maps of task-related experiments \cite{shen2005analysis,sipola2013diffusion}. Shen et al.  \cite{shen2005analysis} employed Diffusion Maps to separate activated from non-activated voxels. Sipola et al.\cite{sipola2013diffusion} used Diffusion Maps with a Gaussian kernel to cluster selected fMRI spatial maps that are derived by ICA. They demonstrated their approach using fMRI recordings acquired from healthy participants listening to a stimulus with a rich musical structure. Other applications of Diffusion Maps in neuroimaging  concern the epileptic-seizure prediction and the identification pre-seizure state in EEG timeseries \cite{lian2015multivariate,duncan2013identifying}.
A review on the intersection between manifold learning methods and the construction of FCN can be found in \cite{siettos2016multiscale} and \cite{richiardi2013machine}.\\
Here, we used MDS, ISOMAP and Diffusion Maps to construct embedded FCN from single-subject ICA analysis of rsfMRI data taken from healthy controls and schizophrenia patients. For our demonstrations, we used the COBRE rsfMRI data that are publicly available and have been used recently in many studies \cite{calhoun2012exploring,mayer2013functional,anderson2013decreased,qureshi2017multimodal}. Based on key global graph-theoretical measures of the embedded graphs, we assessed their classification efficiency using several machine learning algorithms, namely Linear standard Support Vector Machines (LSVM), Radial (Radial basis function kernel) Support Vector Machines (RSVM), k-Nearest Neighbour (k-NN), and Artificial Neural Networks (ANN). We also investigated the performance of two most-commonly used metrics, namely the cross-correlation and the euclidean distance. Our analysis showed that Diffusion Maps outperformed all other methods and lagged cross-correlation proved to be a better choice than the euclidean metric.\\
At this point, we should note, that our study does not aim at the best classification performance by trying to find the best possible pre-processing pipe-line of the raw fMRI data and/or the selection of subjects and/or the selection of the best set of graph-theoretical measures that provide the maximum classification. Yet, we aim at using  state-of-the-art manifold learning methods for the construction of the FCN and compare their classification efficiency using only a few key global theoretical graph-measures and compare the obtained results with those derived by similar studies (see e.g. \cite{anderson2013decreased}) using the same pipe-line for data pre-processing and single-subject ICA. To the best of our knowledge, this paper is the first to perform such a thorough comparative analysis of both linear and nonlinear manifold learning on rsfMRI data. It is also the first study to show how Diffusion Maps can be used for the construction of FCN from rsfMRI, assessing also the efficiency of two basic distance metrics, the cross-correlation-based and the Euclidean distance.

\section{MATERIALS AND METHODS}
\label{sec:1}

\subsection{Pre-processing and signal extraction}
\label{sec:2}
For our demonstrations, we performed a basic pre-processing of the raw fMRI data using SPM as also implemented in other studies (see e.g. \cite{anderson2013decreased}). In particular, for the fMRI data processing, we used FEAT (FMRI Expert Analysis Tool) Version 6.00, part of FSL (FMRIB's Software Library, \url{www.fmrib.ox.ac.uk/fsl}). In particular, the following pre-statistics processing was applied: motion correction using MCFLIRT \cite{jenkinson2002improved}, slice-timing correction using Fourier-space time-series phase-shifting; non-brain removal using BET \cite{smith2002fast}, spatial smoothing using a Gaussian kernel of FWHM 5mm, grand-mean intensity normalization of the entire 4D dataset by a single multiplicative factor. ICA AROMA \cite{pruim2015ica} was also implemented to detect and factor out noise-related independent components (ICs) along with a high-pass temporal filtering at 0.01 Hz (100 seconds) that was applied after ICA AROMA procedure as it is highly recommended \cite{pruim2015ica}.\\
We then proceeded with the decomposition of the pre-processed fMRI data to spatial ICs (for each suject) using the RAICAR methodology \cite{yang2008ranking}. In this way, we found the most reproducible spatial ICs over repeated runs as a solution to the well known problem of the variability of the ICA decomposition \cite{himberg2004validating}. This choice was related to the benchmark fMRIdata per se as there was only a single session per subject with relatively small duration (6 minutes); therefore we wouldn't expect a robust ICA decomposition for all subjects (see also the discussion in \cite{cole2010advances}). Another choice would be to perform group-ICA analysis, but we decided to use single-subject ICA in order to have a common ground with the methodologically similar work presented in \cite{anderson2013decreased}. Group-ICA analysis will be performed in a future study.  

\subsection{Ranking and Averaging ICA by Reproducibility (RAICAR)}
\label{sec:3}

\subsubsection{Independent Component Analysis (ICA)}
\label{sec:4}

ICA is a linear data-driven technique that reduces the high-dimensional fMRI $F(t,x,y,z)$ space in a set of $M$ statistically independent spatial components (ICs). This reduction can be represented as:

\begin{equation}
F(t,x,y,z)=\sum_{i=1}^{M} A_{i}(t) C_{i}(x,y,z),
\end{equation}

where $F(t,x,y,z)$ is the measured BOLD signal, ${A_{i}(t)}$ is the temporal amplitude (the matrix $\mathbf{A}$ containing all temporal amplitudes is known as mixing matrix) and ${C_{i}(x,y,z)}$ is the spatial magnitude of the i-th ICA component. While PCA requires that the principal components are uncorrelated and orthogonal, ICA asks for statistical independence between the ICs. Generally,  ICA algorithms are based either on the minimization of mutual information or the maximization of non-gaussianity among components. 
As discussed in the introduction, most of the standard implementations of ICA, such as the one in MELODIC (Multivariate Exploratory Linear Optimized Decomposition into Independent Components) \cite{beckmann2004probabilistic}, which is part of FSL fMRI analysis software package, share similar gradient based optimization algorithms using an iterative scheme whose initial values are generated randomly, thus making the whole process stochastic. As a consequence, results over repeated runs may differ significantly \cite{himberg2004validating}. A solution to this problem is provided by the so-called ranking and averaging ICA by reproducibility (RAICAR) \cite{yang2008ranking} that we briefly describe in the following section.

\subsubsection{Ranking and averaging ICA by reproducibility (RAICAR)}
\label{sec:5}

The RAICAR methodology developed by Yang et al. \cite{yang2008ranking} was addressed to tackle the problem of the ICs variability by performing $K$ ICA realizations. Thus, RAICAR leads to $K$ ``slightly" different mixing matrices $\mathbf{A}_{1},\mathbf{A}_{2} \dots \mathbf{A}_{K}$ and $K$ different sets of ICs $\mathbf{S}_{1},\mathbf{S}_{2} \dots \mathbf{S}_{K}$. Each realization finds a fixed number $M$ of spatial ICs. Then, a cross realization correlation matrix ($\mathbf{CRCM}$) of size $M\cdot K{\times}M\cdot K$ is constructed and the alignment (ICA produces unaligned components) of ICs of each realization takes place  on the basis of the absolute maximum cross correlation among components. Thus, the cross realization correlation matrix reads:
\[
\mathbf{CRCM} = \begin{bmatrix} 
    \mathbf{R}_{1,1} & \mathbf{R}_{1,2} & \dots & \mathbf{R}_{1,K-1} & \mathbf{R}_{1,K}  \\
    \mathbf{R}_{2,1} &   & \dots & \dots & \mathbf{R}_{2,K} \\
    \vdots &  \dots &  \ddots & \dots & \vdots  \\
    \mathbf{R}_{K-1,1} &  \dots &  \dots &  & \mathbf{R}_{K-1,K} \\
    \mathbf{R}_{K,1} &   \mathbf{R}_{K,2} &   \dots &   \mathbf{R}_{K,K-1}    &  \mathbf{R}_{K,K}
    \end{bmatrix}
\]
$\mathbf{R}_{i,j}$ with $i,j=1,2...K$ are submatrices of size $M{\times}M$ and their elements represent the absolute spatial cross correlation coefficients among components and across realizations. CRCM is a symmetric matrix and its diagonal consists of identity matrices which are ignored for the next steps of the algorithm.\\ 
The procedure starts with the identification of the global maximum of the CRCM, thus finding the matched component based on two realizations. At the next step, the methodology seeks for the highest absolute spatial cross correlation coefficients of the identified component in the remaining realizations factoring out all others. The procedure is repeated $M$ times until $M$ aligned components are found.\\
The next step involves the computation of the reproducibility index for each of the aligned components. This is done by constructing the histogram of the absolute spatial cross correlation coefficients of the upper triangle matrix of the CRCM. This histogram tends to be bimodal, as in general, we expect low spatial cross correlation among most of the ICs and high spatial cross correlation only for a few of them.
A spatial cross correlation threshold is applied with the desired value lying in the valley of the histogram between the two modes \cite{yang2008ranking}. Finally, the reproducibility index is computed for each one of the aligned components. This is done by aggregating the supra-threshold absolute cross correlation coefficients of the CRCM for each of the aligned components.\\
The last step of the algorithm is the ranking and averaging of the aligned components in descending order based on the reproducibility index. The selective averaging is applied so that the components are averaged if and only if, the given aligned component has at least one absolute spatial cross correlation coefficient above the threshold across realizations.\\
After applying RAICAR, the ICs are chosen via a cut-off threshold based on the reproducibility index (of each component) that indicates how consistent is the appearance of an IC across realizations.\\
Here, we have set $K=30$ realizations (same as also in \cite{yang2008ranking}); taking more realizations did not change the outcomes of the analysis. The spatial cross correlation threshold was chosen by localizing the minimum of the histogram of absolute cross correlation coefficients of the CRCM. This threshold was specified separately for each subject. The reproducible  ICs  were determined  by the histogram of reproducibility index over the number of components in descending order. The cut-off threshold was set as the half of the maximum reproducibility index value possible $\frac{K(K-1)}{2}\cdot 0.5$ (this choice is the same with the one used in \cite{yang2008ranking}). This cut-off threshold was set equal for all subjects.\\
Subjects with less than 20 reproducible ICs were excluded from further analysis as this number of components resulted in disconnected graphs. Thus, we ended up with 104 subjects out of which 57 were healthy controls and 47 schizophrenia patients.

\subsection{Construction of Functional Connectivity Networks}
\label{sec:6}

For the construction of FCN, we used all combinations between three manifold learning algorithms, namely MDS, ISOMAP and Diffusion Maps and two widely used metrics, namely the lagged cross-correlation \cite{anderson2013decreased,meszlenyi2017resting,hyde2012cross} and the euclidian distance \cite{sipola2013diffusion,venkataraman2009exploring,goutte1999clustering}.

\subsubsection{Construction of FCN based on the lagged cross-correlation}
\label{sec:7}
For every pair of the associated time courses of the ICs, say $\mathbf{A}_i$ and $\mathbf{A}_{j}$, the cross-correlation function (CCF) over a maximum of three time lags (as in \cite{anderson2013decreased}) reads:
\begin{equation}
CCF(\mathbf{A}_{i},\mathbf{A}_{j},l)=\frac{E[(\mathbf{A}_{i,t+l}-\overline{\mathbf{A}}_{i})(\mathbf{A}_{j,t}-\overline{\mathbf{A}}_{j})]}{\sqrt{E[(\mathbf{A}_{i,t}-\overline{\mathbf{A}}_{i})(\mathbf{A}_{j,t}-\overline{\mathbf{A}}_{j}}]},
\end{equation}

where $l$ is the time lag, and $\overline{\mathbf{A}}_{i}$ is the mean value of the whole time series.\\

For the  construction of the FCN connectivity/ correlation matrices, we used a pseudo-distance measure $d_{c}$ defined as (see also \cite{anderson2013decreased}):

\begin{equation}
d_{c}(\mathbf{A}_{i},\mathbf{A}_{j})=1-\max_{l=0,1,2,3}(|CCF(\mathbf{A}_{i},\mathbf{A}_{j},l)|).
\end{equation}

The resulting dis(similarity) matrices  are fully connected and therefore are hardly comparable between subjects (see the discussion in \cite{anderson2013decreased}). Thus, here as a standard practice, (and in all other algorithms described below), we applied thresholding to the (dis)similarity matrices in order to keep the strongest connections of the derived functional connectivity matrix.
In order to factor out the influence of the variable network density on the computation and comparison of graph-theoretical measures across groups \cite{van_den_Heuvel_2017}, we have implemented the approach of proportional thresholding (PT) \cite{van_den_Heuvel_2017}. In particular, we considered a range of levels of PT from 20\% to 70\% with a step of 2\%. Below the threshold of 20\%, graphs became too fragmented, while thresholds above the 70\% resulted in dense graphs, that mostly included noisy and less significant edges (see also the discussion in \cite{algunaid2018schizophrenic} about this issue). If a graph was still fragmented after thresholding, the giant component was used for further analysis.

\subsubsection{Construction of FCN based on the euclidean distance}
\label{sec:8}

The euclidean distance is used in many studies to assess (dis)similarities between fMRI data points \cite{sipola2013diffusion,venkataraman2009exploring,goutte1999clustering}. For time series associated with the independent spatial maps, $\mathbf{A}_{i}$ and $\mathbf{A}_{j}$, the euclidean distance reads: 
\begin{equation}
L_{2}(\mathbf{A}_{i},\mathbf{A}_{j})=\sqrt{\sum_{t=1}^{T}(A_{i,t}-A_{j,t})^2}.
\end{equation}

For the construction of FCN, PT was applied to the euclidean similarity matrices for each individual over the range of $20$\%- $70$\%.\\

\subsection{Construction of FCN with manifold learning algorithms}
\label{sec:9}

Below we present how MDS, ISOMAP and Diffusion Maps can be exploited to construct (embedded) FCN.

\subsubsection{Construction of FCN with MDS}
\label{sec:10}
Multidimensional Scaling \cite{kruskal1964multidimensional} is a linear method of dimensionality reduction that can be used to find similarities between pairs of objects in a low-dimensional (embedded) space. Given a set of $M$ objects/observables  $\mathbf{x}_1,\mathbf{x}_{2},\dots,\mathbf{x}_{M} \in \mathbf{R}^N$, MDS produces a low-dimensional data representation $\mathbf{y}_{1},\mathbf{y}_{2},\dots,\mathbf{y}_{M}  \in  \mathbf{R}^p, p \ll N$ minimizing the objective function:

\begin{equation}
\sum\limits_{i,j,\ i \neq j } \Big( \|\mathbf{x}_{i}-\mathbf{x}_{j}\|-d(\mathbf{x}_i,\mathbf{x}_j)\Big)^2.
\end{equation}

$d(\mathbf{x}_i,\mathbf{x}_j)$ is the (dis)similarity obtained (by any (dis)similarity measure of choice) between all pairs of points $\mathbf{x}_1,\mathbf{x}_{2},\dots,\mathbf{x}_{M} \in  \mathbf{R}^N$. In our case, the observables $\mathbf{x}_i$ are the amplitudes of the spatial ICs $\mathbf{A}_{i, \ i=1,..M} \in \mathbf{R}^N$. Here, $N=150$ (number of time points).\\
The coordinates of the embedded manifold $\mathbf{y}_{1},\mathbf{y}_{2},\dots,\mathbf{y}_{M}$ are given by:

\begin{equation}
[\mathbf{y}_1,\dots,\mathbf{y}_M]= \mathbf{\Lambda}_{p\times p}\cdot \mathbf{V}^{T}_{p \times M}.
\end{equation}

$\mathbf{\Lambda}_{p\times p}$ contains the square roots of the $p$ largest eigenvalues, and $\mathbf{V}^{T}_{p \times M}$ are the corresponding eigenvectors of the matrix:

\begin{equation}
\mathbf{B}=-\frac{1}{2}\mathbf{H}\mathbf{D}^2 \mathbf{H}^T.
\end{equation}

$\mathbf{H}_{M \times M}$ is the double centering matrix defined as:

\begin{equation}
\mathbf{H}=\mathbf{I}- \frac{1}{M} \mathbf{1} \cdot \mathbf{1}^T, \mathbf{1=[1 1 \dots 1]}_{1\times M}.
\end{equation}

Using MDS, the dimensionality reduction of the original data $\mathbf{X}=\mathbf{x}_{1},\mathbf{x}_{2},\dots,\mathbf{x}_{M} \in \mathbf{R}^N$ yields the embedding of $\mathbf{Y}={\mathbf{y}_{1},\mathbf{y}_{2},\dots,\mathbf{y}_{M}  \in \mathbf{R}^p}$, $ p \ll N$.\\

Here, for the construction of the embedded FCN, we produced distance matrices (using either lagged cross-correlation or the euclidean distance) $\mathbf{D_{Y}}$ of size $M \times M$.\\ 

For the implementation of the MDS algorithm, we used the ``cmdscale" function contained in the software package``Stats" in the R free Software Environment \cite{team2014r}.

\subsubsection{Construction of FCN using ISOMAP}
\label{sec:11}
ISOMAP is a non-linear  manifold learning algorithm that given a set of $M$ objects/observables $\mathbf{x}_{1},\mathbf{x}_{2},\dots,\mathbf{x}_{M} \in \mathbf{R}^N$ produces a low-dimensional data representation $\mathbf{y}_{1},\mathbf{y}_{2},\dots,\mathbf{y}_{M} \in \mathbf{R}^p$, $p \ll N$ minimizing the objective function:

\begin{equation}
\sum\limits_{i,j, \ i \neq j } \Big( \ d_G(\mathbf{x}_i,\mathbf{x}_j)- d(\mathbf{x}_{i},\mathbf{x}_{j})\Big)^2,
\end{equation}

where $d_G(\mathbf{x}_i,\mathbf{x}_j)$ is the shortest path (geodesic distance)  and $d(\mathbf{x}_i,\mathbf{x}_j)$ is the (dis)similarity obtained (by any (dis)similarity measure of choice) between all pairs of points $\mathbf{x}_{1},\mathbf{x}_{2},\dots,\mathbf{x}_{M} \in \mathbf{R}^N$.\\
In our case, the observables $\mathbf{x}_i$ are the amplitudes of the spatial ICs $\mathbf{A}_{i, \ i=1,..M} \in \mathbf{R}^N$.\\
The above minimization problem is solved as follows: \cite{tenenbaum2000global}:\\
\begin{itemize}
    \item Construct a graph $\mathbf{G}=(V,E)$, where the vertices $V$ are the ICs $\mathbf{A}_i$; its links $E$ are created by using either the $k$-nearest neighbors algorithm or a fixed distance between nodes, known as the $\epsilon$ distance.
    For example, a link between two ICs is created if $d_{i,j}\equiv d(\mathbf{A}_i,\mathbf{A}_j)< \epsilon \ , \ \forall \ i \neq j$. Here, we used the $k$ nearest neighbours algorithm with $k=3,4,5,6$ \cite{tenenbaum2000global}).
    Set the weight $w_{i,j}$ of the link (if any) between $\mathbf{A}_i,\mathbf{A}_j$ as $w_{i,j}= \frac{1}{d(\mathbf{A}_i,\mathbf{A}_j)}$. If there is not a link set: $w_{i,j}=0$.\\
    \item Approximate the  embedded manifold by estimating the shortest path (geodesic distance) $d_G (\mathbf{A}_i,\mathbf{A}_j)$ for each pair of nodes based on the distances $d_{i,j}$; 
    this step can be implemented for example using the Dijkstra algorithm \cite{dijkstra1959note}. This procedure results to a matrix, $\mathbf{D_G}$ with elements the shortest paths:
    \begin{equation}
    D_{G_{ij}} \equiv d_G (\mathbf{A}_i,\mathbf{A}_j)=min \big\{ d_{i,j},d_{i,k}+d_{k,j} \big\}, \,k=1,2,\dots,M \quad k \neq i,j.
    \end{equation}
    \\
    \item Estimate the coordinates of the low-dimensional (embedded) manifold $\mathbf{y}_1,\mathbf{y}_2,\dots,\mathbf{y}_M$  exploiting the MDS algorithm \cite{kruskal1964multidimensional} on the geodesic distance matrix $\mathbf{D_G}$. 
\end{itemize}
Here, for the implementation of the ISOMAP algorithm, we used the package ``vegan" \cite{oksanen2007vegan} contained in the R free Software Environment \cite{team2014r}. 
\\





\subsubsection{Diffusion maps}
\label{sec:12}
Diffusion Maps \cite{coifman2006diffusion} is a non-linear manifold learning algorithm that given a set of $M$ objects/observables $\mathbf{X}=\mathbf{x}_{1},\mathbf{x}_{2},\dots,\mathbf{x}_{M} \in \mathbf{R}^N$ produces a low-dimensional representation $\mathbf{Y}={\mathbf{y}_{1},\mathbf{y}_{2},\dots,\mathbf{y}_{M}}  \in \mathbf{R}^p$, $p \ll N$, addressing the diffusion distance among data points as the preserved metric\cite{nadler2006diffusion}.
The embedding of the data in the low-dimensional space is obtained by the projections on the eigenvectors of a normalized graph Laplacian \cite{belkin2003laplacian}.  The Diffusion Maps algorithm can be described in a nutshell in the following steps: 

\begin{itemize}
    \item  Construction of the affinity matrix $\mathbf{W}_{M \times M}$, here $M$ is the number of ICs for each subject. The elements $W_{ij}$ represent the weighted edges connecting nodes $i$ and $j$ using the so-called heat kernel:
\begin{equation}
W_{i,j}= exp(- \frac{d(\mathbf{x}_{i},\mathbf{x}_{j})^2}{\sigma} ),
\end{equation}

where $\mathbf{x}_i$ is a $N$-dimensional point (here, N=150), $d(\mathbf{x}_i,\mathbf{x}_j)$ are the (dis)similarities obtained (by any dissimilarity measure of choice) between all pairs of points $\mathbf{x}_{1},\mathbf{x}_{2},\dots,\mathbf{x}_{M} \in \mathbf{R}^N$ and $\sigma$ is an appropriately chosen parameter which can be physically described as a scale parameter of the heat kernel \cite{coifman2006diffusion}. The heat kernel $\mathbf{W}$ satisfies two important properties, the one of symmetry and the other of positive semi-definite matrix. The latter property is crucial and allows the interpretation of weights as scaled probabilities of ``jumping" from one node to another.\par

The parameter $\sigma$ of the neighborhood
size is data-dependent and here, it was determined by finding the linear region in the sum of all weights in $\mathbf{W}$, say $S_w$, using different values of $\sigma$ \cite{singer2009detecting,sipola2013diffusion}. $S_w$ is calculated through the formula:

\begin{equation}\label{sumof}
S_w=\sum\limits_{i}^M\sum\limits_{j}^M{W_{ij}},
\end{equation}

In order to use a single value of $\sigma$ for all participants, we computed a super-distribution of the sum of weights across subjects (taking the median value of the distributions) using different values of $\sigma$. 
Thus, we considered values of $\sigma$ lying in the linear region of the super-distribution. After inspection, every value of $\sigma$ was lying in the linear region of every single subject's logarithmic plot.\\

    \item Formulation of the diagonal $M \times M$ normalization matrix $\mathbf{K}$ along with the diffusion matrix $\mathbf{P}$:
\begin{equation}
K_{ii}=\sum_{j=1}^{M} W_{ij},
\end{equation}

\begin{equation}\label{oldm}
\mathbf{P=K}^{-1}\mathbf{W}. 
\end{equation}

Each element of the symmetric and normalized diffusion matrix $\mathbf{P}$ reflects the connectivity between two data points $\mathbf{x}_{1}$ and $\mathbf{x}_{2}$. As an analogy, this connectivity can be seen as the probability of ``jumping" from one point to another in a random walk process. Consequently, raising $\mathbf{P}$ to a power of $t$ can be thought as a diffusion process. As the number of $t$ increases, paths with low probability tend to zero, while the connectivity between paths with high probability remains high enough, thus governing the diffusion process \cite{coifman2006diffusion}. Thus, the algorithm of Diffusion Maps preserves the diffusion distance among points in a low-dimensional euclidean space. The diffusion distance is closely related to the diffusion matrix $\mathbf{P}$ and for two distinct points $\mathbf{x}_{i}$, $\mathbf{x}_{j}$ and for specific time instance $t$ is defined as \cite{de2008introduction}:
\begin{equation}
D_{t}(\mathbf{x}_{i},\mathbf{x}_{j})=\sum_{m}|P_{im}^{t}-P_{mj}^{t}|^2.
\end{equation}
Unlike geodesic distance, the diffusion distance is robust to noise perturbations, as it sums
over all possible paths (of $t$ steps) between points \cite{coifman2006diffusion}.\\
    \item Construction of the conjugate matrix 
\begin{equation}\label{newm}
\mathbf{ \overline{P}}= \mathbf{K}^{1/2}\mathbf{P}\mathbf{K}^{-1/2},
\end{equation}
substituting Eq.(\ref{oldm}) to Eq.(\ref{newm}) we get  
\begin{equation}
\mathbf{\overline{P}}= \mathbf{K}^{-1/2}\mathbf{W}\mathbf{K}^{-1/2}.
\end{equation}
This is the so-called Graph Laplacian matrix \cite{belkin2003laplacian}. The matrix $\mathbf{\ P}$ is adjoint to the symmetric matrix $\mathbf{\overline{P}}$. Thus, $\mathbf{\ P}$ and $\mathbf{\overline{P}}$ share the same eigenvalues \cite{nadler2008diffusion}.\\ 
 \item Singular Value Decomposition (SVD) of $\mathbf{\overline{P}}$ yields
 
\begin{equation}
\mathbf{\overline{P}}=\mathbf{U}\mathbf{\ \Lambda} \mathbf{U}^{*},
\end{equation}
where $\mathbf{\Lambda}$ is a diagonal matrix containing the $M$ eigenvalues of $\mathbf{P}$ and $\mathbf{ U}$ the eigenvectors of $\mathbf{\overline{P}}$. The eigenvectors  $\mathbf{V}$  of $\mathbf{P}$ can be bow found by \cite{nadler2008diffusion}:
\begin{equation}
\mathbf{\ V}=\mathbf{K}^{1/2}\mathbf{U}.
\end{equation}

    \item By taking out the trivial eigenvalue $\lambda=1$ of the matrix $\mathbf{\Lambda}$ and the corresponding eigenvector contained in $\mathbf{V}$, the coordinates of the low dimensional embedded manifold $\mathbf{y}_{1},\mathbf{y}_{2},\dots,\mathbf{y}_{M}$ are given by:
     \begin{equation}
[\mathbf{y}_1,\dots,\mathbf{y}_M]= \mathbf{\Lambda}_{p\times p}\cdot \mathbf{V}^{T}_{p \times M},
\end{equation}
where $\mathbf{\Lambda}_{p\times p}$ contains the $p$ largest eigenvalues, and $\mathbf{V}^{T}_{p \times M}$ are the corresponding eigenvectors of the diffusion matrix $\mathbf{P}$. 
\end{itemize}
For the implementation of the above algorithm we used the package ``diffusionMap" \cite{richards2014diffusionmap} contained in the R free Software Environment \cite{team2014r}.

\subsubsection{Choice of the embedded dimension}
\label{Embdim}
Here, the embedded dimension was determined via the eigenspectrum of the final decomposition for every dimensionality reduction/Manifold learning algorithm. If there is a gap between the first few  larger eigenvalues  of the final decomposition and the rest of the eigenspectrum, these few eigenmodes capture most of the distance differences between data points and are able to represent and uncover intrinsic properties of the data structure \cite{nadler2008diffusion,strange2014open,saul2006spectral}. In order to determine the embedded dimension for the methods described above, we considered the following steps:
We sorted the eigenvalues in decreasing order eg. $\lambda_{1} \geq \lambda_{2} \geq \lambda_{3} \dots \geq \lambda_{M}$ (for diffusion maps $\lambda_{1}$ is discarded). Then, for each subject, we calculated the consequent pairwise differences $\lambda_{1}-\lambda_{2}$, $\lambda_{2}-\lambda_{3}$, ... , $\lambda_{M-1}-\lambda_{M}$.  A gap between the mean (over all subjects) differences determines from which point and further, the relative contribution of another eigendimension is redundant (this having a small contribution to the reconstruction of the embedded FCN). Thus, here for our computations, we considered low-dimensional embeddings of $p=(2,3,4,5)$ dimensions for each one of the manifold learning methods described above.

\subsection{Graph-theoretical measures} 
\label{sec:13}
Here, we analyzed the topological properties of the binary FCN graphs on the basis of three major graph measures, namely, the average path length, the global clustering coefficient, and the median degree, which are the most frequently used in neuroscience \cite{stam2007graph}.
In particular, given a graph $\mathbf{G}=(V,E)$ with $g_{ij}$ representing the link (0: unconnected or 1: connected) from node $i$ to  node $j$ and $k_i= \sum^{N_V}_{j=1}g_{ij}$ the degree of node i, the graph measures are computed as follows:\\
a) The average path length is defined by: $L= \frac{1}{N_V(N_V-1)}\sum_{i \neq j}D_{G_{ij}}$, i.e. is the average number of steps along the shortest paths $D_{G_{ij}}$ for all possible pairs of the network nodes. This is a measure of the efficiency of information or mass transport on a network between all possible pairs of nodes.\\
b) The global clustering coefficient is defined by: $C_g= \frac{\sum_{} t_c}{\sum_{} t}$, where $t$ is a triplet and $t_c$ is a closed triplet. A triplet of a graph consists of  three nodes that are connected by either to open (i.e open triplet) or closed (i.e closed triplet) ties. In general, this measure indicates how random or structured a graph is (in our case, in terms of functional segregation).\\
c) The median degree $M_{k}$ is the median value of the degree distribution of $\mathbf{G}$.
This measure reflects how well connected is the ``median" network node in terms of the number of links that coincide with it.\\
An extensive review of definitions and meaning of the above and other graph-theoretical measures with respect to brain functional networks can be found in \cite{rubinov2010complex,stam2007graph}.\\
The computations for the graph analysis were performed with the ``igraph" \cite{csardi2006igraph} package contained in the free software environment R \cite{team2014r}.

\subsection{Classification Algorithms}
\label{sec:14}
Based on the three key graph measures, the classification was assessed using machine learning algorithms, namely Linear Support Vector Machines(LSVM), Radial Support Vector Machines (RSVM), Artificial Neural Networks (ANN) and k-NN  classification (for a  brief description of the above algorithms and their parameters see in the Appendix). The classification algorithms were trained, validated and tested using a 10-fold cross-validation scheme which was repeated 100 times. Thus, we separated the data in ten separate sub-samples; nine of them were used as training sets and one of them was used for validation. This process was repeated 10 times leaving out each time a different sub-sample which served as a validation set. The whole procedure was repeated 100 times. The overall classification rate was determined via the computation of the average classification rate over all the repetitions of the 10-fold cross validation for each model.\\
The average confusion matrix (over all repetitions of the 10-fold cross validation) was also computed for each classification model. The confusion matrix is a  $2\times2$ (in the case of binary classification) square matrix  containing all true positives $TP$, false positives $FP$, true negatives $TN$ and false negatives $FN$. Here, we considered as positives $P$ the schizophrenia cases and as negatives $N$ the healthy control cases. Sensitivity (also called the True Positive Rate) and Specificity (also called the True Negative Rate) are basic statistical measures for the assessment of  binary classifications. The sensitivity $TPR$  is given by $TPR= \frac{TP}{TP+FN}$, while the specificity $TNR$ is given by $TNR= \frac{TN}{TN+FP}$. Here, sensitivity characterizes the ability of the classifier to correctly identify a schizophrenic subject, while specificity is the ability of the classifier to correctly identify a healthy subject.\\ 
Here, we used the algorithms contained in the package ``caret" \cite{kuhn2008building} contained in the R free software environment \cite{team2014r}.

\section{RESULTS} 
\label{sec:15}

\subsection{Classification performance using the lagged cross-correlation metric}
\label{sec:16}

In table \ref{fig:table 1}, we present the best classification accuracy, along with the corresponding sensitivity and specificity rate for each manifold learning algorithm, threshold and classifier when using the lagged cross-correlation metric. At the end of the table \ref{fig:table 1}, we provide also the results obtained by the ``conventional" (thresholded) cross-correlation matrix. 

\begin{table}[!htb]
\caption{Best classification rates over all manifold and machine learning methods using the lagged cross-correlation pseudo-distance measure $d_c$ (see \ref{sec:7}); the embedded dimension (Emb. Dim) is also shown for each method along with the corresponding best threshold point (Thres), Classifier, Accuracy (Acc), Sensitivity (Sens) and Specificity (Spec) rate.}
\renewcommand{\arraystretch}{1.5}  

 \resizebox{0.92\textwidth}{!}{\begin{minipage}{\textwidth}
 \centering
\begin{tabular}{ c c c c c c c } 
\hline  
\textbf{\textit{Method}} & \textbf{\textit{Emb. Dim}}&\textbf{\textit{Thres}}&   \textbf{\textit{Classifier}} & \textbf{\textit{Acc}} $\pm$ \textbf{\textit{SD}}   & \textbf{\textit{Sens}}& \textbf{\textit{Spec}}\\  
\hline 
\textit{MDS} &3 &0.3 & RSVM  & 68.4$\pm$1.3 \% & 51.1\% & 77.8\%\\ 
\textit{} &3  &0.36  &  LSVM &58.6$\pm$1.7 \% &28.8\% &78.9\% \\
\textit{} &3  &0.3  &  k-NN &63.2$\pm$2.1 \% &51\% &68.8\% \\
\textit{} &3  &0.3  &  ANN &63.6$\pm$2.6 \% &50.7\% &69.8\% \\
\hline 
\textit{Isomap} &2 &0.24 & RSVM  &74.4$\pm$1.9 \%  & 69.4\% & 73.1\% \\
\textit{} &2  &0.28  &  LSVM  &64.4$\pm$1.7 \% &55.8\% &67.1\% \\
\textit{} &2  &0.24  &  k-NN &71$\pm$2.1 \% &62.9\% &72.7\% \\
\textit{} &2  &0.24  &  ANN &68.8$\pm$3 \% &63.1\% &68.6\% \\
\hline 
\textit{Diffusion Maps} &4 &0.52 &RSVM &79.3$\pm$1.2 \% & 74.1\% & 77.9\% \\
\textit{} &4  &0.56  &  LSVM  &72.2$\pm$1.7 \% &66.6\% &71.8\% \\
\textit{} &4  &0.52  &  k-NN &74.7$\pm$1.4 \% &72.3\% &71.5\% \\
\textit{} &4  &0.52  &  ANN &78.6$\pm$2 \% &74.1\% &76.8\% \\
\hline   
\textit{Correlation Matrix} &- &0.52 &RSVM  &69.5$\pm$1.5 \% & 77.1\% & 58.3\% \\
\textit{} &-  &0.52  &  LSVM  &71$\pm$1.5 \% &75.8\% &61.9\% \\
\textit{} &-  &0.34  &  k-NN &67.2$\pm$2.2 \% &57.8\% &70.1\% \\
\textit{} &-  &0.52  &  ANN &68.8$\pm$1.6 \% &68.3\% &64.3\% \\
 \\  

\end{tabular} 
 \end{minipage}}
\label{fig:table 1}  
\end{table}



\begin{figure*}[!htb]
\centering
\includegraphics[width=0.6\linewidth]{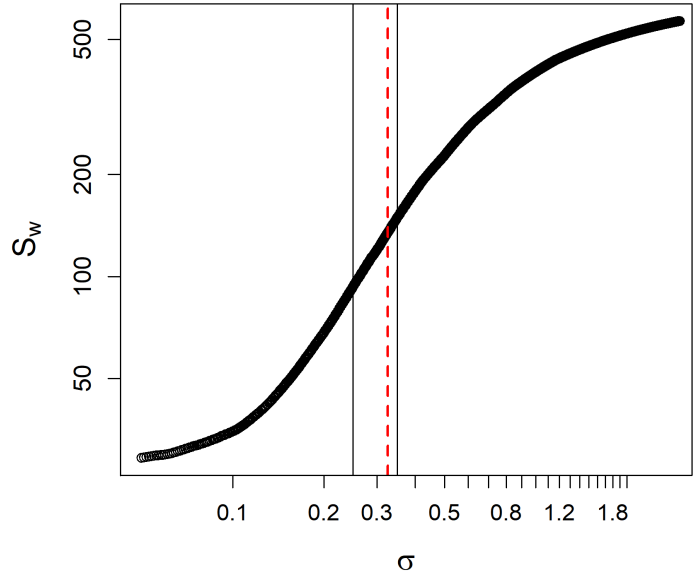}
\caption{\textbf{Super-distribution of all subjects of the sum of the weights (see the equation in \ref{sumof}) (e.g. by taking the median value of the distribution across subjects per value of $\sigma$) using different values of $\sigma$ parameter. The red dashed vertical line shows the optimal $\sigma$ that was found to be $\sigma$= 0.325. The other two vertical black lines depict the linear zone in which we investigated values of $\sigma$.}}
\label{fig:figure2}
\end{figure*}

Fig. \ref{fig:figure2} shows the super-distribution of the sum of weights of all subjects against different values of $\sigma$ used for construction of the FCN using the Diffusion Maps algorithm; the red dotted vertical line shows the optimal $\sigma$ (here, $\sigma$ = 0.325) while the black vertical lines bound the linear region ($\sigma \in (0.25,0.35) $). The results were robust to different choices of the time step of the Diffusion Maps algorithm, namely $t=0,1,2$. The investigation of the optimal $\sigma$ was performed with respect to the best classification accuracy over all classifiers.\\


\begin{figure*}[ht!]
\centering
\includegraphics[width=1\linewidth]{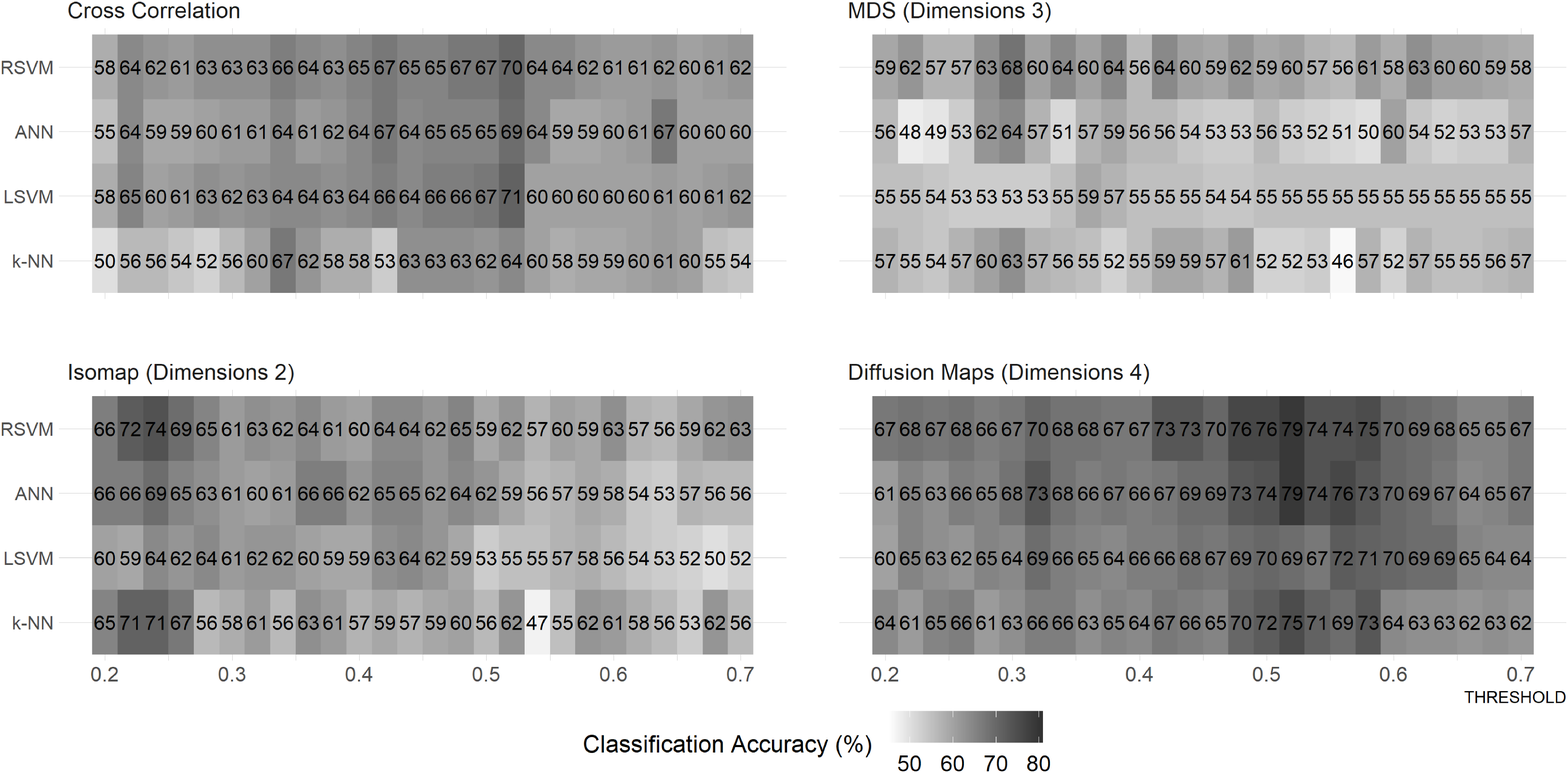}
\caption{\textbf{Overall classification performance for all thresholds (from 20\%-70\% of the strongest edges with 2\% as pace) and classifiers. The metric used is the lagged cross-correlation based pseudo-distance measure $d_c$ (see \ref{sec:7}).}}
\label{fig:figure3}
\end{figure*}

Fig.\ref{fig:figure3} depicts the best classification rates for MDS, ISOMAP, Diffusion Maps and lagged-cross correlation matrix for different classifiers are reported. Diffusion Maps provided the best classification accuracy (79.3\% for SVM with an RBF kernel and 52\% PT), thus appearing more robust over a wide range of PT. With respect to the maximum classification accuracy obtained by Diffusion Maps, results were robust over a wide range of values of $\sigma \in (0.28,0.35)$. 

\begin{figure*}[ht!]
\centering
\includegraphics[width=0.7\linewidth]{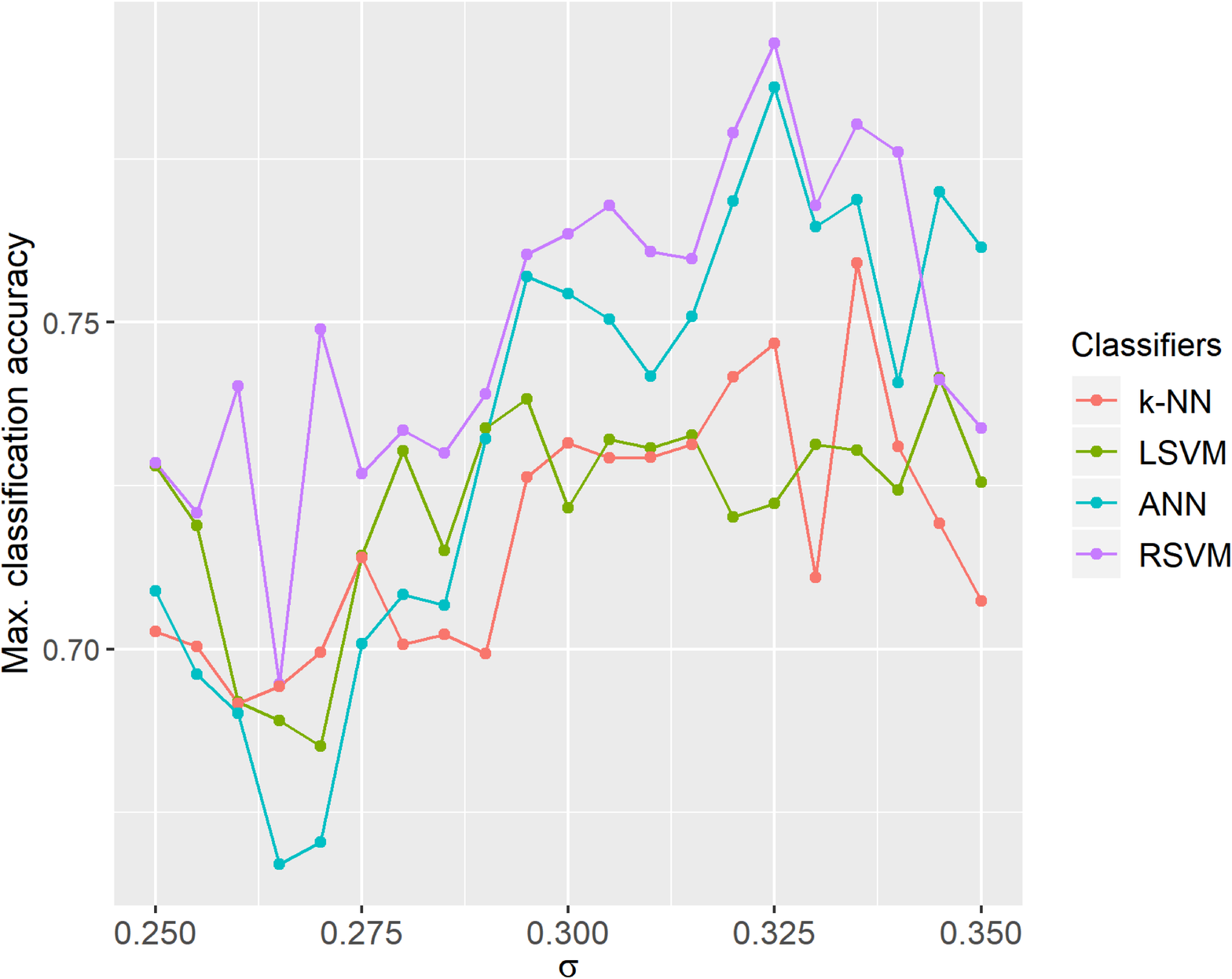}
\caption{\textbf{Classification performance of Diffusion Maps for different values of $\sigma$ and classifiers. The metric used is the pseudo-distance measure $d_c$ (see \ref{sec:7}) based on lagged cross-correlation.}}
\label{fig:figure4}
\end{figure*}

Fig. \ref{fig:figure4} depicts the classification performance obtained with Diffusion Maps for different values of $\sigma$.\\

Isomap performed better for low thresholds. Overall, the performance of Isomap was more or less the same with the cross-correlation matrix with the best classification rate being slightly higher for Isomap (74.4\% for SVM with an RBF kernel and 24\% PT). Its performance was however sensitive to the choice of the number $k$ of nearest neighbors; with $k=5$ we got a 74.4\% classification accuracy for SVM with an RBF kernel and 24\% PT) while for $k=6$ we got a 71\% classification accuracy for ANN and 20\% PT), thus providing the best performance for ISOMAP. MDS was  outperformed by both ISOMAP and Diffusion Maps as well as the  cross correlation matrix; the classification performance of MDS was overall relatively poor.  

\begin{figure*}[ht!]
\centering
\includegraphics[width=0.8\linewidth]{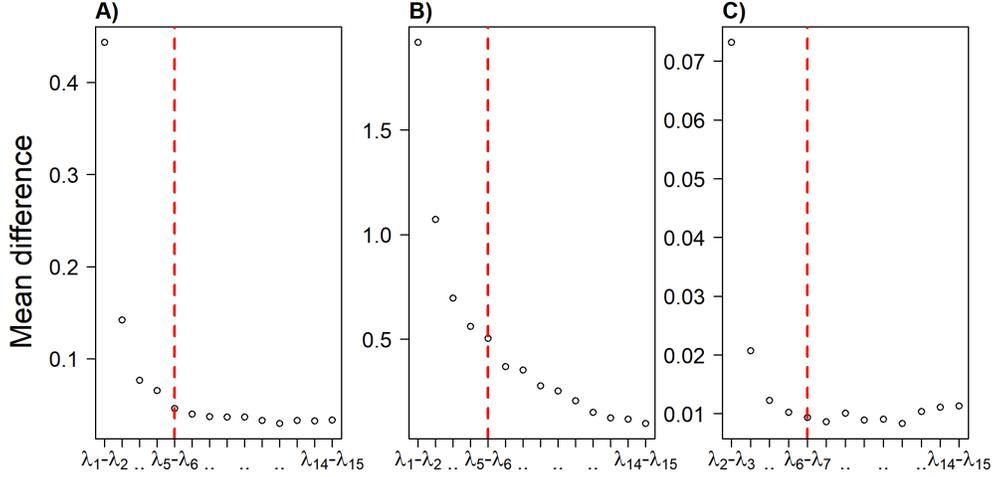}
\caption{\textbf{Mean Differences of the first larger 15 eigenvalues (see \ref{Embdim}) for all manifold learning algorithms with the cross correlation metric. A) MDS (see  \ref{sec:10}), B) ISOMAP (see  \ref{sec:11}) and C) Diffusion Maps (see  \ref{sec:12}), based on the optimal parameters. The red dashed vertical line marks the maximum number of dimensions considered in this study (i.e. five dimensions).}}
\label{fig:last}
\end{figure*}

Fig. \ref{fig:last} shows characteristic  eigenspectrums of the final decomposition for MDS, ISOMAL and Diffusion Maps. As it is shown, there are three gaps: the first gap appears between the first eigenvalue and the rest of the spectrum, the second gap between the first two eigenvalues and the rest of the spectrum, and a third gap appears between the first four-five eigenvalues and the rest of the spectrum.

\subsection{Classification performance using the euclidean distance}
\label{sec:17}

The same analysis was performed for the euclidean distance. 
The best classification rates using the euclidean distance for all manifold learning methods and classifiers are presented in table \ref{fig:table 2}.\\

\begin{table}
\caption{Best classification rates over all manifold learning methods and classifiers with the use of the euclidean distance $L_{2}$ (see \ref{sec:8}); the embedded dimension (Emb. Dim), threshold point (Thres), classifier,  Accuracy (Acc), Sensitivity (Sens) and Specificity (Spec) rates.}
\renewcommand{\arraystretch}{1.5}  
\resizebox{0.92\textwidth}{!}{\begin{minipage}{\textwidth}
\centering
\begin{tabular}{ c c c c c c c }
\hline  
\textbf{\textit{Method}} & \textbf{\textit{Emb. Dim}}&\textbf{\textit{Thres}}&   \textbf{\textit{Classifier}} & \textbf{\textit{Acc}} $\pm$ \textbf{\textit{SD}}  & \textbf{\textit{Sens}}& \textbf{\textit{Spec}}\\  
\hline  
\textit{MDS} &3 &0.54 & RSVM  & 72.3$\pm$1.7 \% & 52.8\% & 83.4\%\\ 
\textit{} &3  &0.36  &  LSVM &57.9$\pm$2.5 \% &26.6\% &79.6\% \\
\textit{} &3  &0.26  &  k-NN &65.7$\pm$2.2 \% &59.6\% &66.1\% \\
\textit{} &3  &0.44  &  ANN &64.9$\pm$2.6 \% &53.7\% &69.5\% \\
\hline 
\textit{Isomap} &3 &0.68 & RSVM    &72.9$\pm$2 \%  & 55.8\% & 81.9\% \\
\textit{} &3  &0.42  &  LSVM  &54.6$\pm$0.1 \% &0\% &95.9\% \\
\textit{} &3  &0.68  &  k-NN &65$\pm$2.3 \% &47.7\% &74.7\% \\
\textit{} &3  &0.66  &  ANN &68.2$\pm$1.8 \% &47.4\% &80.4\% \\
\hline 
\textit{Diffusion Maps} &5 &0.66 &RSVM &68.8$\pm$2.2 \% & 57.7\% & 73.1\% \\
\textit{} &5  &0.52  &  LSVM &62.9$\pm$1.9 \% &56.7\% &63.5\% \\
\textit{} &5  &0.26  &  k-NN &65.1$\pm$2.5 \% &63.3\% &61.7\% \\
\textit{} &5  &0.5  &  ANN &63.9$\pm$2.6 \% &71.4\% &53.2\% \\
\hline   
\textit{Euclidean Matrix} &- &0.64 &RSVM &71.6$\pm$1.7 \% & 60.1\% & 76.1\% \\
\textit{} &-  &0.58  &  LSVM &59.2$\pm$2.5  \% &62.5\% &52.3\% \\
\textit{} &-  &0.64  &  k-NN &72$\pm$2.2 \% &67.5\% &70.6\% \\
\textit{} &-  &0.70  &  ANN & 62.1$\pm$2.6 \% &60.4\% &59.1\% \\
 \\  
\end{tabular}  
 \end{minipage}}
\label{fig:table 2}  
\end{table}



Overall, the  classification rates with the euclidean distance were more noisy compared to the ones computed with the lagged cross-correlation distance.\\
For a wide range of PT, the classification rates obtained were under 60\% and sometimes at the level of a random or a completely biased classifier (towards the larger class, here, the healthy controls). In our case, a random classifier resulted to a 50.5\% classification accuracy, while for a strict majority rule for imbalanced datasets, a completely biased (and thus useless) classification model results to a 54.8 \% classification rate.\\

\begin{figure*}[ht!]
\centering
\includegraphics[width=1\linewidth]{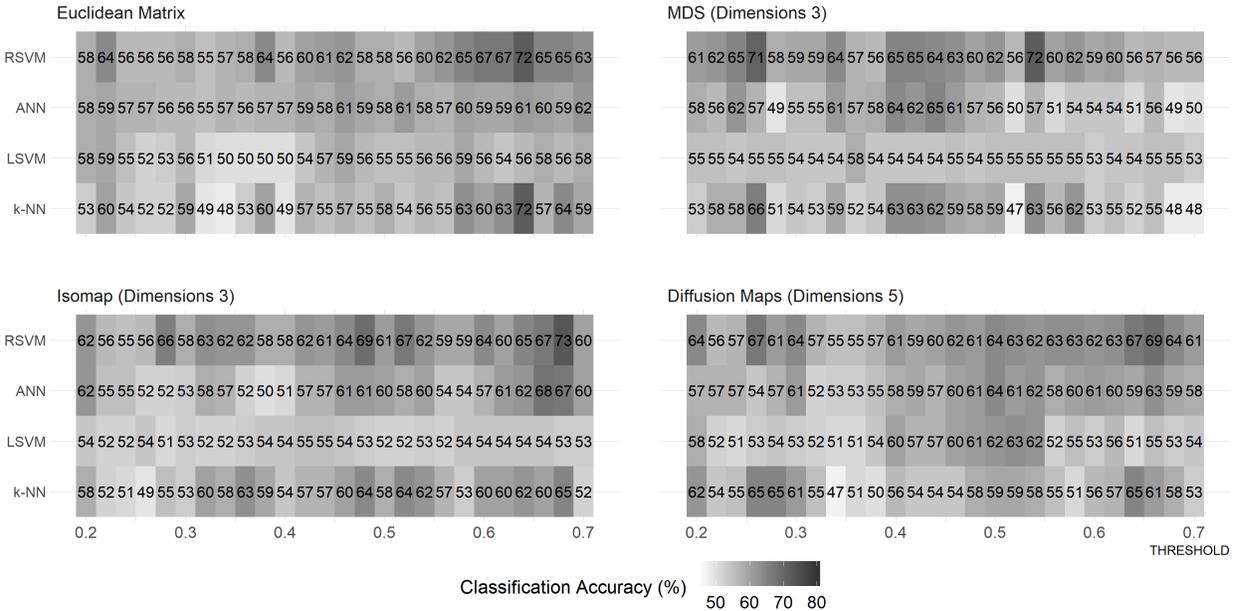}
\caption{\textbf{Classification performance using the euclidean distance  (see \ref{sec:8}) for all thresholds (from 20\%-70\% of the strongest edges with 2\% as pace) and classifiers.}}  
\label{fig:figure6}
\end{figure*}

Fig. \ref{fig:figure6} depicts the accuracy of all methods across all thresholds using all different classifiers.

\begin{figure*}[ht!]
\centering
\includegraphics[width=0.8\linewidth]{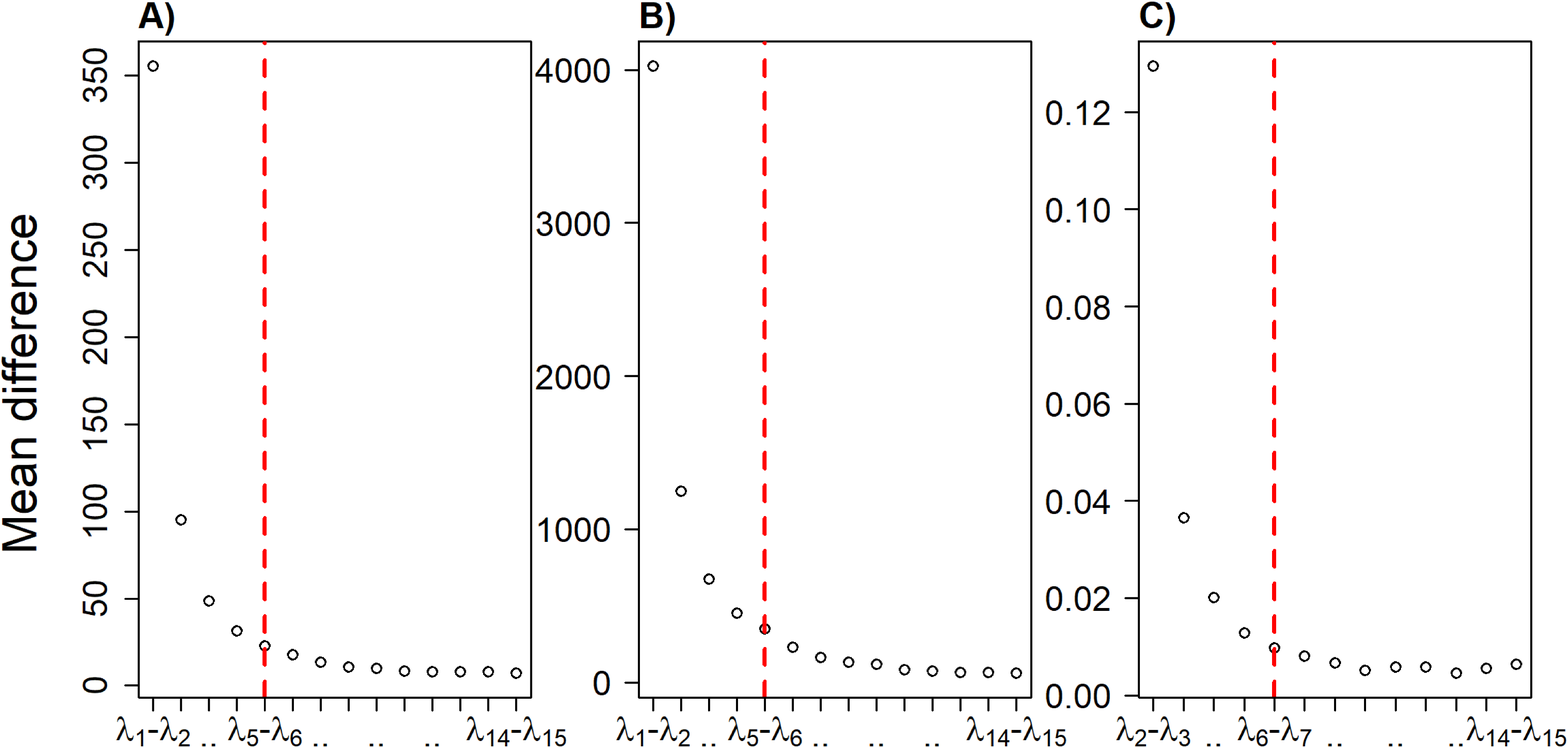}
\caption{\textbf{Mean Differences of the first larger 15 eigenvalues (see \ref{Embdim}) for all manifold learning algorithms with the euclidean metric. A) MDS (see  \ref{sec:10}), B) ISOMAP (see  \ref{sec:11}) and C) Diffusion Maps (see  \ref{sec:12}) using the optimal parameters. The red dashed vertical line shows marks the maximum number of dimensions considered in this study (i.e. five dimensions).}}
\label{fig:lasteucl}
\end{figure*}

\begin{figure*}[ht!]
\centering
\includegraphics[width=0.9\linewidth]{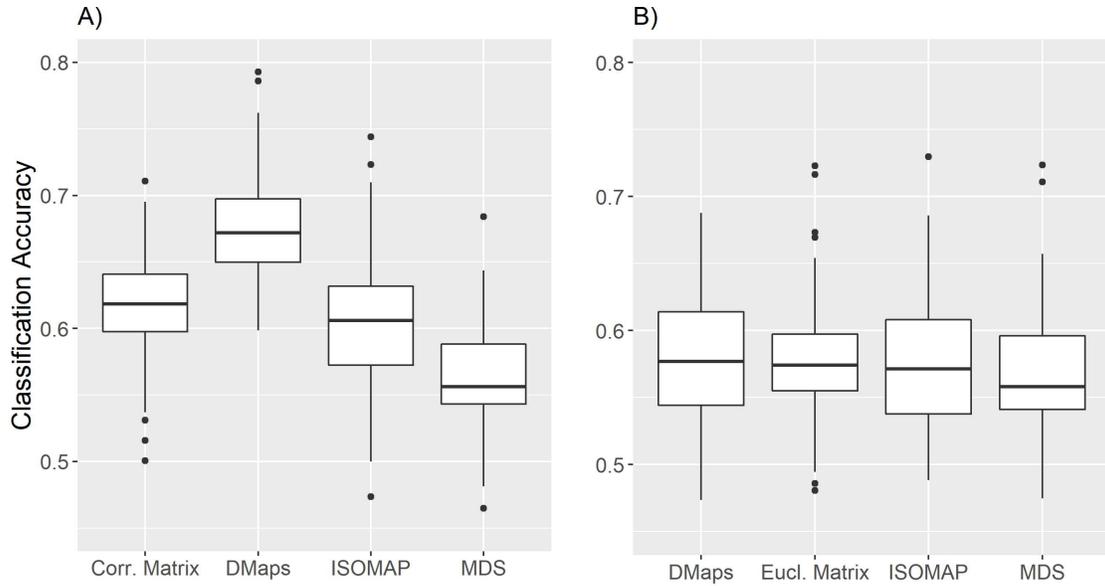}
\caption{\textbf{Boxplots of classification rates over all classifiers and thresholds, using the A) lagged cross-correlation pseudo-distance $d_c$ (see \ref{sec:7}).  B) the euclidean distance $L_{2}$ (see \ref{sec:8}). The labels at the bottom of each panel  correspond to the method used for the construction of the FCN: the Cross Correlation matrix (Corr. Matrix), Diffusion Maps (DMaps), ISOMAP, MDS. The black horizontal lines mark the median values of the distributions.}}   

\label{fig:figure7}
\end{figure*}

In this case, the best classification rate was obtained with Isomap (72.9\% for RSVM and 68\% PT). Despite the fact that for certain thresholds, the maximum classification rates obtained with the euclidean distance were enough high, the performance of the methods was generally poorer compared to the ones computed with the lagged cross-correlation distance for all the range of PT.\\
Characteristic  eigenspectrums for all manifold learning algorithms utilizing the euclidean distance are shown in Fig. \ref{fig:lasteucl}.\\
Finally, in Fig. \ref{fig:figure7} are given the boxplots of the classification rates among metrics used. 
Overall, the lagged cross-correlation distance resulted in higher classification rates compared to the euclidean distance with the exception of MDS. Except from some specific thresholds, MDS in both cases performed poorly.

\section{DISCUSSION} 
\label{sec:18}
In this study, we constructed embedded FCN from  rsfMRI data using linear and non-linear manifold learning techniques. Based on the graph theoretical measures of the constructed FCN we then used machine learning algorithms for classification purposes. We also compared the performance of two widely used metrics, namely the lagged cross-correlation and the euclidian metric. For our illustrations, we used a publicly available dataset of resting fMRI recordings taken from healthy and patients with schizophrenia. This is the first study that performs such a systematic comparative analysis between various manifold learning algorithms, machine learning algorithms and metrics. To the best of our knowledge, it is also the first study that shows how Diffusion Maps can be exploited to construct FCN from fMRI data.\\
At this point we should note that our intention was not to try to obtain the best possible classification performance by ``optimising" the pre-processing of the raw fMRI data and/or by trying to find the best set of graph-theoretical measures. Instead we used a very basic pre-processing of the raw data (as the one performed in \cite{anderson2013decreased}), while classification was based only on three key graph-theoretical measures in neuroscience \cite{stam2007graph}, namely, the average path length, the global clustering coefficient and the median degree  of the embedded binary networks. Based on the above, our best reported accuracy was 79.3 \% obtained with Diffusion Maps and lagged cross-correlation (using 10 fold cross validation repeated 100 times).\\
For the same benchmark fMRI dataset Anderson and Cohen \cite{anderson2013decreased} used ISOMAP for the construction of embedded low-dimensional FCN for the classification between controls and schizophrenia patients. ROIs were acquired using single subject ICA and functional connectivity was accessed using the lagged cross-correlation distance. The analysis revealed differences in small world properties among groups and 13 graph theoretic features led to a reported 65\% accuracy rate. Algunaid et al. \cite{algunaid2018schizophrenic} reported a 95\% accuracy (with FSV feature selection method and 82.5\% with a Welch's t-test) testing more than 360 graph-based features. AAL was used for signal extraction and ROI selection, while an SVM-based classifier was used along with 10 fold cross validation scheme to evaluate its performance.\\
An important outcome of our analysis is that the Diffusion Maps result in more robust results and higher classification accuracy and that the lagged cross-correlation distance outperforms the euclidean distance which is used traditionally to construct connectivity matrices from fMRI data.\\
In a future work, we aim at analysing the local-graph theoretical properties, thus matching the derived RAICAR ICs with known resting state networks and also perform group ICA analysis.

\section{Data availability}
The COBRE dataset is publicly available at \url{http://fcon_1000.projects.nitrc.org/indi/retro/cobre.html}. For our analysis, we used the ``R" software subroutines as described in \ref{sec:1} and \ref{sec:19}.

\section{Author Contributions}

Conceptualization:Constantinos Siettos; Methodology: Constantinos Siettos; Formal analysis and investigation: Ioannis Gallos; Data acquisition and processing: Ioannis Gallos and Evangelos Galaris; Writing- original draft preparation: Ioannis Gallos; Writing - review and editing: Constantinos Siettos and Ioannis Gallos; Supervision:Constantinos Siettos

\section{APPENDIX} 
\label{sec:19}
 
 \ For classification we used Linear Support Vector Machines (LSVM), Radial (Radial basis function kernel) Support Vector Machines (RSVM), one hidden layer Artificial Neural Networks (ANN) and k-NN classifier(k-NN). All classifiers  were trained and evaluated via repeated 10-fold cross-validation scheme repeated 100 times. 
 For the classification we used the three key graph theoretical measures as described in subsection \ref{sec:13} in Materials and Methods.
 Training and classification were implemented using algorithms contained in package ``caret" \cite{kuhn2008building} publicly available in R free software environment \cite{team2014r}.

 \subsection{Support Vector Machines (SVM)}
 Support vectors machines (SVM) aim at finding the optimal separating plane or hyperplane in the feature space among groups. In particular, for a set of points $(\mathbf{x}_i,y_i)_{i=1,2..N}$, where $N$ is the number of subjects, $\mathbf{x}_i\in \mathbf{R}^d$ contains $d$ attributes/features selected for subject $i$ and $y_i \in (-1,1)$ the subject's class (here, either healthy or patient), SVM attempts to find the optimal plane or hyperplane that divides the two classes by maximizing the margin of separation. Any hyperplane with a given set of points $\mathbf{x}_i$ can be modelled as $\mathbf{w}\cdot \mathbf{x}_i+b=0$ where $\mathbf{w}$ represent the weights of features $\mathbf{x}_i$. Parallel hyperplanes can be described as $\mathbf{w}\cdot \mathbf{x}_i+b \geq 1$ if  $y_i=1$  and $\mathbf{w}\cdot \mathbf{x}_i+b \leq -1$ if  $y_i=-1$. The optimization problem then aims at maximizing the margin between hyperplanes $\frac{2}{\|\mathbf{w}\|}$ such that for every $(y_i)_{i=1,2..N}, \, y_i\cdot (\mathbf{w}\cdot \mathbf{x}_i+b) \geq 1$.
One can take advantage of a regularization parameter $C$ indicating the penalty of error $z_i$ that gives a trade-off between misclassifications and the width of the separating margin. This leads to the final optimization problem, which minimizes $\frac{\|\mathbf{w}\|^2}{2}+ C\cdot \sum_{i}z_i$ subject to $y_i\cdot (\mathbf{w} \cdot\mathbf{x}_i+b) \geq 1-z_i$, \, $i=1,2..N$.\\
Based on the idea that the data maybe better separable in a higher dimensional space, SVM may utilize  a kernel function to map $\mathbf{x}_i\in \mathbf{R}^d$ to $\phi (\mathbf{x}_i)\in \mathbf{R}^D$, $D>d$. In our study, besides standard Linear SVM (LSVM), we also used Radial SVM (RSVM) making use of the Radial basis functions kernel given by $K(\mathbf{x}_i,\mathbf{x}_j)=exp(-\frac{\|\mathbf{x}_i-\mathbf{x}_j\|^{2}}{2\cdot \gamma^2})$, where $\gamma$ is the kernel's scale parameter.

\subsection{k-Nearest Neighbours classifier(k-NN)}
k-nearest neighbours algorithm is one of the simplest classification/machine learning algorithms. Given $(\mathbf{x}_i,y_i)_ {i=1,2..N}$, where $N$ is the number of subjects, $\mathbf{x}_i\in \mathbf{R}^d$ contains $d$ attributes/features selected for subject $i$ and $y_i$ the subject's class (here, either healthy or patient), k-NN utilizes euclidean distance in the feature space to perform a voting system among $\kappa$ closest neighbours. In this manner, each point is classified as ``control", if the number of``control" neighbours is greater than the number of ``patient" neighbours and inversely. The number $\kappa$ of closest neighbours is a parameter of choice that plays a crucial role in method's performance. In this study, it is important to note that we chose odd values of $\kappa$ (i.e how many neighbours we take into consideration) in order not to have to break possible ties in the voting system among neighbours

\subsection{Artificial Neural Networks (ANN)}
In this study, we also used feed-forward  Artificial Neural Networks (ANN) consisting of one hidden layer. The input units were three as the number of the features considered for classification. We have chosen one hidden layer consisting from 1 to 5 neurons along with a bias term. The activation function used for all neurons was the logistic transfer function \cite{ripley2007pattern}. The output was one node (reflecting simple binary classification control/patient). The training procedure of the model was done via back-propagation \cite{hecht1992theory} using a 10-fold cross validation scheme repeated 100 times. Finally a weight decay parameter $a$  (regularization parameter) was used to prevent over-fitting and improve generalization \cite{krogh1992simple} of the final model. For the implementation of the ANN we used the ``nnet" software package \cite{ripley2011nnet} publicly available in R free software environment \cite{team2014r}.

\subsection{Parameters tested for each classifier}

We tuned the parameters of the algorithms via grid search.\\
For the SVM:
$C= (0.1,0.25,0.5,0.75,1,2.5,5,7.5,10,25,50,75,100,250,500,750,1000)$\\
$\gamma= (0.001,0.01,0.1,0.25,0.5,0.75,1,2.5,5,7.5,10,25,50,75,100,250,500,750,1000)$.\\
For the k-NN classifier:
$\kappa= (1,3,5,7,9)$.\\
For the ANN:
number of neurons in the hidden layer $p= (1,2,3,4,5)$,\\
decay level $a= (0.0001,0.001,0.01,0.025,0.05,0.075,0.1)$.




\end{document}